\documentclass[a4]{article}

\usepackage{graphicx}
\usepackage{dcolumn}
\usepackage{multicol} 
\usepackage{bm}
\usepackage{a4wide}

\begin{document}

\begin{center}
\LARGE{
Extensive collection of femtoliter pad secretion droplets
in beetle {\it Leptinotarsa decemlineata} allows \\nanoliter microrheology}

\vspace{0.5cm}

\Large{B\'ereng\`ere Abou$^{1}$, Cyprien Gay$^{1}$, Bastien
  Laurent$^{1}$, Olivier Cardoso$^{1}$,\\ Dagmar Voigt$^{2}$, Henrik
  Peisker$^{3}$, Stanislav Gorb$^{2,3}$}

\vspace{0.5cm}

\noindent \normalsize{
$^{1}$Laboratoire Mati\`ere et Syst\`emes
Complexes (MSC), UMR CNRS 7057 \& Universit\'e Paris Diderot, Paris, France,\\
$^{2}$Evolutionary Biomaterials Group, Max-Planck-Institut für
Metallforschung,\\ Heisenbergstr. 3, 70569 Stuttgart, Germany,\\
$^{3}$Department of Functional Morphology and
  Biomechanics, Zoological Institute of the University of Kiel, Am
  Botanischen Garten 1-9, 24098 Kiel, Germany
}
\end{center}
\date{\today}

\begin{abstract}

Pads of beetles are covered with long, deformable setae, each ending
in a micrometric terminal plate coated with secretory fluid. It was
recently shown that the layer of the
pad secretion covering the terminal plates is responsible for the
generation of strong attractive forces. However, less is known
about the fluid itself because it is
produced in extremely small quantity. We here present
a first experimental investigation of the rheological properties of
the pad secretion in the Colorado potato beetle {\it Leptinotarsa
decemlineata}. Because the secretion is
produced in an extremely small amount at the level of the terminal
plate, we first develop a procedure based on capillary effects to
collect the secretion. We then manage to incorporate micrometric
beads, initially in the form of a dry powder, and record their thermal motion
to determine the mechanical properties of the surrounding medium. 
We achieve such a quantitative measurement
within the collected volume,
much smaller than the $1\,{\rm \mu}$l sample volume
usually required for this technique.
Surprisingly, the beetle secretion was found to behave as a
purely viscous liquid, of high viscosity. This suggests that 
no specific complex fluid behaviour is
needed during beetle locomotion. 
We build a
scenario for the contact formation between the spatula at the setal
tip and a substrate, during the insect walk. 
We show that the attachment dynamics of the insect pad computed from the high measured viscosity
is in good agreement with observed insect pace.
We finally discuss the consequences of the secretion viscosity
on the insect adhesion.
\end{abstract}

\section{Introduction}

Among few other animal groups, insects possess a fascinating
ability to walk on smooth vertical surfaces and even on ceilings. Such
ability is robust, fault tolerant and resistant to
contamination. Insects can stick well to both hydrophobic and
hydrophilic surfaces and detach in a very fast manner. The comparative
data show that the evolution of attachment mechanisms in insects has
developed along two distinctly different mechanisms: smooth flexible
pads and hairy (setose, fibrillar) surfaces \cite{Gorb2001,
Beutel2001, Beutel2006}. In cockroaches, grasshoppers and bugs, pads
are soft deformable structures with relatively smooth surface, whereas
in flies, beetles and earwigs, they are covered with long deformable
setae. Because of the flexibility of the material in both mechanisms --
soft attachment pads or fine surface microstructures -- the possible contact area with the wide
range of substrate profiles is maximized despite the surface roughness
of the substrates, and high proximity between
contacting surfaces can ensure strong attachment forces between pad and
substrata.

Hairy attachment systems are typical for evolutionary younger and
successful insect groups, such as flies and beetles and have a huge
diversity of forms. There are several geometrical effects, such as
multiple contact formation, high aspect ratio of single contact
structures, peeling prevention using spatula-like tips of single
contact elements that are responsible for the generation of a strong
pull-off force. These effects found in
attachment devices of insects are an important source of information
for further development of biomimetic patterned adhesives. 
One may speculate about the different physical mechanisms
that are able to generate sufficient
adhesion despite variation in the physico-chemical properties of the
surface (hydrophobic, hydrophilic), surface profile (rough, smooth)
and environmental condition (dry, wet). 
The theoretical background pertaining to these physical effects has been
intensively theoretically discussed in several recent publications
\cite{Arzt2003, Persson2003, PerssonandGorb2003, Chung2005, Gao2005}.

Even though hairy attachment systems have been under investigation for
more than 300 years~\cite{Leeuwenhoek}, 
the attachment mechanism of animals walking on
smooth walls or ceiling is still under debate. Different hypotheses
for attachment, such as sticky fluid, microsuckers or the action of
electrostatic forces, have been proposed
\cite{dewitz1884,simmermacher1884,gillett1932}. It is now well known
that hairy attachment pads of reduviid bugs \cite{Edwards1970}, flies
\cite{Bauchhenss1979, Walker1985} and beetles \cite{Ishii1987,
Kosaki1996, Eisner2000} secrete fluid into the contact area, whereas
others, such as spiders or geckos, do not.
Based on experimental data,
adhesion has been recently attributed to molecular interactions and
capillary attractive forces mediated by secretion (reduviid bugs,
flies, beetles)~\cite{Stork1980} or Van der Waals interactions
(spiders, geckos)~\cite{autumn2000,autumn2002}.

\section{Motivation}

In hairy wet attachment systems, the role of the secretion still remains
unclear. It has already been shown that the presence of the
fluid is required for generating adhesion in insect adhesive pads.
For example, as shown in 1970, attachment was impaired when
hairy pads of the bug {\it Rhodnius prolixus} were treated with
organic solvents \cite{Edwards1970}. 
In 1980, experiments with beetles on various substrates
have also strongly suggested that cohesive forces, 
surface tension and molecular adhesion, 
mediated by pad secretion, may be involved 
in the mechanism of attachment~\cite{Stork1980}. 
It was infered from experiments~\cite{Wallentin1999,Gorb2001}
that the secretion induces a viscous resistance to
detachment of pad from substrate and a capillary attraction which
applies both before and during detachment (static an dynamic
process).
Later, on the scale of individual seta terminal plates,
experiments conducted on fly {\it Calliphora vicina}
using multiple local force-volume AFM measurements
have demonstrated that the adhesion between the AFM tip and the seta
(used to estimate the local seta adhesive properties)
is about two times stronger in the
center of the terminal plate, where thickness of the fluid is higher,
than on its border~\cite{Langer2004}:
adhesion was shown to strongly decrease as
the volume of the secretion decreases, indicating that a layer of
pad secretion, covering the terminal plates, is crucial for generation
of the strong attractive force. 

However, by measuring attachment forces on a smooth glass using a
centrifuge technique, recent experiments
demonstrate two apparently contradictory results~\cite{Dreschler2006}:
{\em (i)} both friction and adhesion of insects pads 
on smooth surface are greater when less secretion is present,
and {\em (ii)} the adhesive force on rough substrates
decreases when the pad secretion is depleted
through multiple consecutive pull-off motions. 
From a physiological point of view, these results
seem to suggest that the most important function of the secretion 
is to provide attachment on rough substrates.
Indeed, most of the substrates
encountered by insects are not smooth but rough. 
From a physical point of view,
these results closely reproduce two complementary requirements
well known for adhesive substances, whether natural or manufactured:
they must {\em (a)} establish a good contact with the substrate
even in the presence of surface roughness
and {\em (b)} resist separation, in other words dissipate
a significant amount of energy during separation~\cite{PHYSTOD}.

Usual pressure-sensitive adhesive materials
such as those included in adhesive tapes or stickers,
which are solids,
manage to establish a good contact (point {\em (a)})
as a result of their high deformability,
as recognized by Dahlquist more than 40 years ago~\cite{Dahlquist1966}.
{\em A fortiori}, a liquid secretion 
is suitable for establishing an intimate contact with a rough substrate
and helps maximizing the contact area.

Concerning the strong resistance during separation (point {\em (b)}),
it implies that the secretion be a very dissipative material.
In other words, the loss modulus $G^{\prime\prime}$
in the relevant frequency range corresponds 
to a significant amount of dissipation.
This property may hinder the ability
of the secretion to establish a good initial contact
within a timescale compatible with the insect walk.
For this reason, it is crucial to determine
the rheology of the secretory material.
That issue is the central goal of the present work.

\section{About known relevant secretion properties}

Secretion is known to contain a non-volatile, lipid-like substance in
diverse insects, but in some other groups, such as flies and ants, it
is a two-phasic microemulsion presumably containing water-soluble and
lipid-soluble fractions~\cite{Gorb2001, Federle2002, Votsch2002}. 
Concerning its mechanical properties,
up to now, measurements were interpreted 
assuming that secretion was purely viscous.
The secretion dissipation has been shown 
to be large,
although a quantitative evaluation of its rheological properties 
could not be conducted because the exact geometry of droplets, 
capillary bridges, and contact areas were unknown~\cite{Wallentin1999,Langer2004}.
Observing the dewetting velocity of ants' secretion droplets
through Interference Reflection Microscopy
were interpreted in terms of viscosity
and provided an estimate of $40$ to $150$~mPa.s~\cite{Federle2002}.
In this experiment, because the dewetting velocity is constant,
the mechanical properties of the material are probed at a fixed frequency,
and no inference can be made about the response at other frequencies
and hence about the possibly visco-elastic properties.

\section{Our mechanical measurements}

In the present paper, we quantitatively evaluate
the rheological properties of the attachment pad secretion
in the beetle {\it L. decemlineata} for the first time
and we relate them to the attachment dynamics 
of the insect.

\begin{figure}
\includegraphics[angle=0,width=8.0cm]{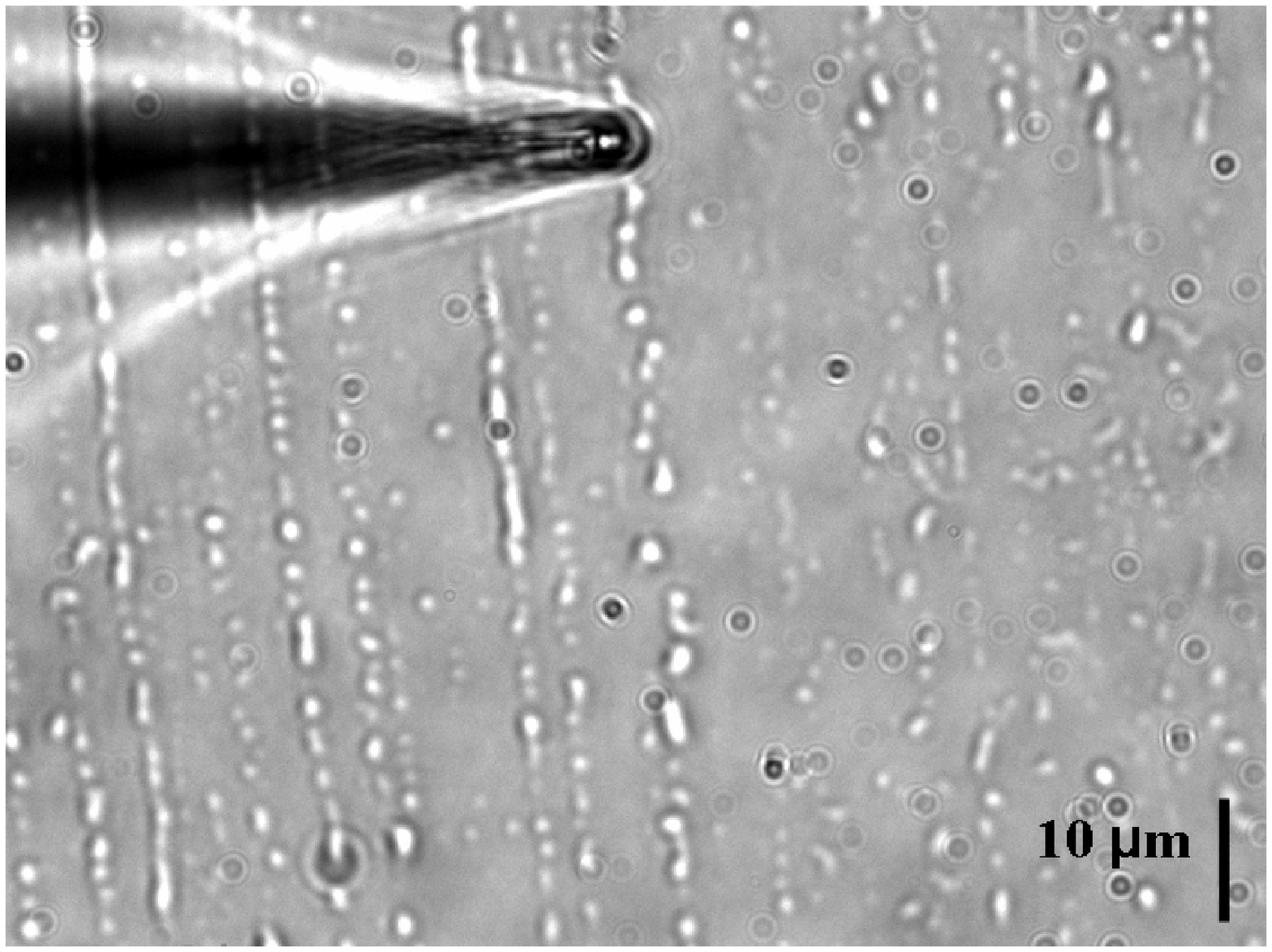}
\caption{Secretion droplets from adults beetles {\it
    L. decemlineata} on a glass slide.
The beetles' legs are rubbed against the slide, 
which results in a strong shear that is able to deposit the secretion on
    the surface. The
  typical size of the droplets is a few microns. A home-made
  microneedle, mounted on a three-axes micromanipulator, was used to
    draw up the secretion droplets spread on the glass slide. The tip
    was moved onto the slide surface to collect the largest amount of
    secretion. Due to capillary
  effects, the rise of the liquid takes place
  spontaneously inside the
  capillary tube.
}
\label{aspiration}
\end{figure}

The essential difficulty of such a rheological measurement
resides in the {\em extremely tiny} amount of secreted fluid: 
it totally excludes the use of standard rheology techniques
which require at least a few milliliters of material.
Microrheology, by contrast, typically requires 
less than 1 microliter of sample
and is thus suitable for performing rheological measurements
in situations where the available volume of material is a limiting factor.
This issue is particulartly vivid with the insect secretion:
the typical volume of a secretory droplet 
of the Colorado potato beetle {\it Leptinotarsa
decemlineata}, collected on a glass slide and shown in Fig.~\ref{aspiration},
is $1\ \mu$m$^3$, which corresponds to $10^{-9}\ \mu$l. 
At first sight, in order to obtain the required volume of $1\ \mu$l,
it would be necessary to collect a discouragingly large number
of droplets: typically $10^9$ droplets!
In the present work, as we shall see,
we achieve a reliable microrheological measurement
with a much smaller sample volume,
namely $10^{-4}\,\mu{\rm l}$ ($0.1$~nl).

The microrheology technique involves using
micrometric beads to measure the relation between stress (probe force)
and deformation (probe position) in materials, at the microscopic
scale. The driving force applied on the probes is thermal, with an
energy scale corresponding to $kT$. 
Measurements of the particles' mean-squared displacement (MSD) 
give access to the possibly visco-elastic properties of the fluid material.
Because the driving force is
small, only the linear viscoelastic response of the material is
probed. From the stress-deformation relation, the visco-elastic
properties of the surrounding medium can be derived~\cite{mason,waigh}.

In the following, we first propose a procedure to collect the
secretion, in a sufficiently large amount, by using capillary effects,
taking place in a home-made microneedle. 
We then describe how micrometric probes were 
immersed in the collected volume, and their Brownian motion
recorded. 
We then validate our measurements performed in the secretion
(statistical accuracy and geometry)
with additional tests in a calibrated Newtonian oil.
We finally suggest a scenario for the contact formation 
between a single seta and the substrate in such a fluid. 
Using this scenario, together with the rheological data,
we then estimate the attachment dynamics of the setae,
and therefore of the insect pads.

\section{Collecting the secretion using capillarity} 
A home-made microneedle, with a few microns tip, 
was used to draw up the secretion
droplets (1 to 10 $\mu$m in diameter) spread on a glass slide (Figure
\ref{aspiration}). The procedure used to collect the secretion from insects is described in the Appendix. The microneedles were pulled from
thin-wall borosilicate glass capillaries with 1 mm of the outer diameter and
0.78 mm of inner diameter (Harvard Apparatus, France) with a Narishige
PB-7 double-stage puller (Narishige Instruments, Tokyo, Japan). By
adjusting the puller settings, microneedles with tip of a few microns
were produced. The microneedle was then mounted on a three axes piezo
micro-manipulator Burleigh, and attached to an inverted
microscope (Leica DMI3000).

\begin{figure}
 \includegraphics[angle=0,width=8.0cm]{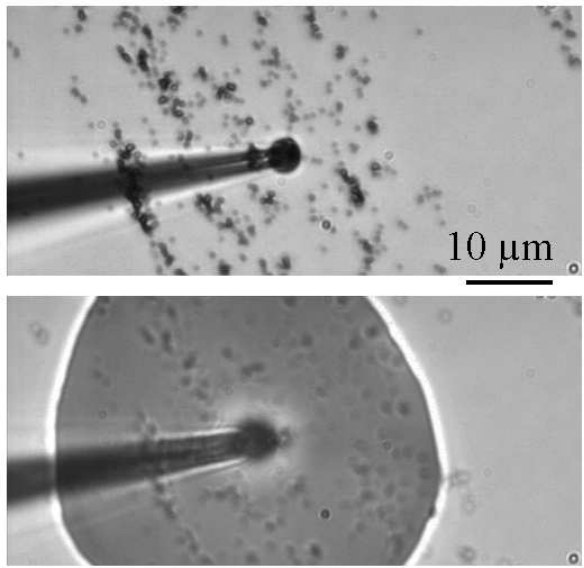}
\caption{The collected volume is ejected on dry $0.74 \, \mu{\rm m}$ in
  diameter Melamine beads (Top) before the ejection (the dry beads
  have been deposited on a glass slide) and  
(Bottom) during the ejection. The final volume of collected secretion 
is a drop of $100 \,\mu{\rm m}$ in diameter, and $30\, \mu{\rm m}$ in height.}
\label{ejection}
\end{figure}

Due to capillary effects, the rise of the wetting secretion takes
place spontaneously in the tube as the tip is maintained in contact
with the support and the droplet. The microneedle tip was moved onto
the slide surface in order to collect the largest possibe amount of
secretions. Due to its high flexibility, it could be maintained in
contact with the slide without breaking, thus optimizing the collection of the secretion. The microneedle tip was connected to a syringe, allowing
us to apply a positive or negative pressure to the microneedle. 
Spontaneous suction -- due to surface tension -- 
could possibly be assisted by applying slight negative pressure to
the microneedle. After an entire day ($\sim 8$ hours) collecting the
secretions, the final volume of fluid represented a drop 
which was about $100\, \mu{\rm m}$ in diameter, and $30\, \mu{\rm m}$ in height 
(Figure~\ref{ejection}).

\section{Brownian motion measurements} 

\begin{figure}
 \includegraphics[angle=0,width=8.0cm]{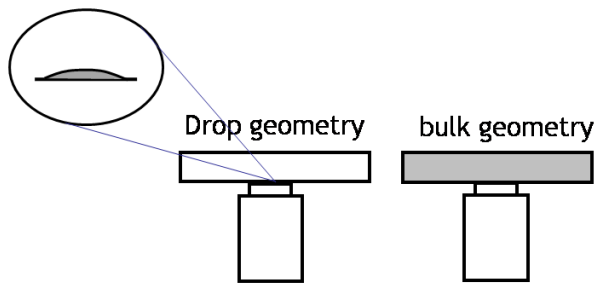}
\caption{Scheme of 'bulk' and 'drop' geometries used to perform microrheological
  measurements. The bulk geometry ($ 1 {\rm cm} \times 1 {\rm cm}
  \times 100  \, \mu {\rm m}$) is the usual geometry to perform
  microrheological measurements. The 'drop' geometry was used with
 the secretion in the present work, due to the extremely 
  small amount of fluid available. The drop size was $100  \, \mu {\rm
  m}$ in diameter and $30  \, \mu {\rm
  m}$ in height. The 'drop' geometry was shown to provide the
  same viscosity results as the
  usual bulk geometry, which shows that it is reliable.}
\label{geometry}
\end{figure}

Dry powder of melamine
beads (Acil, France) was deposited on a clean glass slide. The
collected secretion volume was then ejected on the dry beads, which
were $0.740 \pm 0.005~\mu$m in diameter, by applying a positive pressure to the
microneedle (Figure \ref{ejection}). During the ejection, the
microneedle was slightly pulled away, in order to avoid secretion to
rise along the outer side of the microneedle tube. After ejection, the microneedle tip
was moved along the glass slide to unstick beads from the surface. The secretory fluid
was then drawn up again and ejected several times in order to mix
beads with secretion. The final mixture was then ejected in a square chamber ($10
\times 10 \,{\rm mm}^2$) made of a microscope plate and a cover-slip
separated by a thin adhesive spacer ($100 \,\mu$m thickness). The
container was sealed to avoid contamination or possible evaporation of
the secretion drop (Figure \ref{geometry}). 

The fluctuating motion of the tracer beads,
immersed in the secretory fluid, was recorded with a fast CCD camera
(Kodak motion) 
mounted on an inverted Leica microscope, with an oil immersion
objective (63 X). The fast camera was typically sampling at 125 Hz,
during $8$ s. For reliable
analysis of the Brownian motion, particular attention was brought to
record the motion far from the rigid wall imposed by the glass slide
and the drop edges. A home-made image analysis software allowed us to
track the beads positions $x(t)$ and $y(t)$ close to the focus plane
of the objective (see Appendix for more details). For each bead, the time-averaged mean-squared
displacement $\langle\Delta r^2(t)\rangle_{t'}=
\langle[x(t'+t)-x(t')]^2 +[y(t'+t)-y(t')]^2 \rangle_{t'}=2
\langle\Delta x^2(t)\rangle_{t'}$ is calculated, improving the
statistical accuracy.  To preserve a reliable statistics, the data of
the mean-squared displacement were kept in the range below $ t < 2$
s~\cite{abou08}. The quantity $\langle\Delta r^2(t)\rangle_{t'}$ was then averaged
over several beads (about $20$), and identified as the
ensemble-averaged mean-squared displacement. The resolution on the
bead position, determined with a sub-pixel accuracy in the image
analysis detection, was about $0.3$ pixel corresponding to $\delta =
30$ nm. This resolution determines the lowest accessible MSD in these experiments,
and corresponds to about $2.10^{-3} \,\mu{\rm m}^2$.

\begin{figure}
 \includegraphics[angle=0,width=7.5cm]{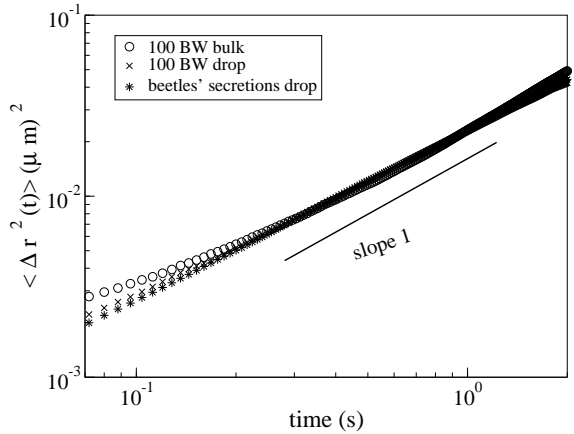}
\caption{Mean-squared displacement of $0.74 \, \mu {\rm m}$ Melamine beads immersed
  in: $(*)$ the beetle secretion, 
$(o)$ the calibrated 100 BW Newtonian fluid in the bulk geometry, and
  $(x)$ 100 BW in the drop geometry, 
as a function of time. In all cases, the fluctuating motion of the
  tracers is purely diffusive, 
characterized by a linear dependency of the MSD with time. The
  secretion was found to behave as a 
purely viscous fluid on the time scales investigated, of viscosity
  about one hundred time the water viscosity.}
\label{results}
\end{figure}

Figure~\ref{results} shows the ensemble-averaged mean-squared
displacement $ \langle\Delta r^2(t)\rangle_{E}$ (MSD) of the tracer
beads, as a function of the lag time $t$. As can be seen, a diffusive
behaviour of the tracers, characterized by a linear dependency of the
MSD with time, is measured. This linear dependency implies that the
beetles' pad secretion simply behave in a purely viscous manner in the
investigated time
window. The secretory fluid viscosity corresponds to $110 \pm 5 $
mPa.s at the room temperature $T = 21 \pm 1 ^{\tiny{\rm o}}$C. This
value corresponds to about one hundred times the water viscosity. 

In order to test the reliability of the Brownian motion measurements
in the 'drop' geometry used for secretion, microrheological tests were
also performed in a calibrated Newtonian fluid (100 BW,
ZMK-ANALYTIC-GmbH), in two different geometries. The first one -- bulk
geometry -- corresponds to the usual geometry used in microrheology
($10 {\rm mm} \times 10 {\rm mm} \times 100 \,\mu$m), while the second
one -- drop geometry -- corresponds to a drop, approximately
$100\,\mu$m in diameter and $30\,\mu$m in height, spread on a glass
slide (Figure \ref{geometry}). In both experiments, the Brownian
motion of the tracers was recorded, far from the glass slide walls and
the drop edges.  Also, in both experiments, the MSD was averaged on
about the same beads number, of the order of $20$.  The experimental
results were found to be in good agreement in both geometries, in the
time range investigated $7.10^{-2} < t < 2.0 $ s (Figure
\ref{results}). In both cases, the MSD increases linearly with time,
indicating a purely viscous behaviour, characterized by a viscosity
of the order of $100$ mPa.s, in excellent agreement with standard
rheological measurements in the calibrated oil. The 'drop geometry',
as well as the statistical accuracy, were then confirmed to be
reliable in our microrheological measurements.

\section{Insect pad attachment dynamics}

\begin{figure}
\includegraphics[angle=0,width=8.0cm]{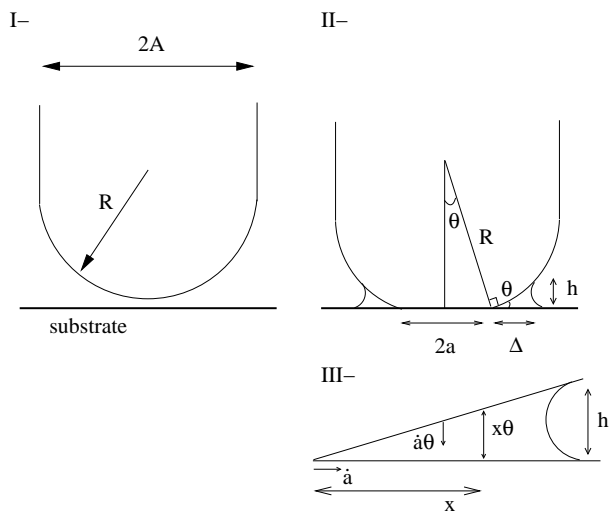}
\caption{Geometry of the setal tip. The seta
consists in a tip of diameter $2A$ and curvature $1/R$ located at the
end of a long fiber (I). The tip of the seta is covered by a thin
film of viscous secretion. When the seta contacts substrate, the contact
region expands laterally.
(II). The meniscus (outer diameter $2 (a + \Delta) $) 
still represents a small portion of the tip surface ($a\ll A$). 
(III). As the contact broadens ($\dot{a}>0$),
the tip surface approaches the substrate at velocity $\dot{a}\,\theta$.
The viscous dissipation in the meniscus 
then determines the velocity $\dot{a}>0$.}
\label{spatula}
\end{figure}

Let us now focus on the modelling of the attachment dynamics of the
insect in the presence of secretion. The spatula at the setal tip is
considered to be a portion of a hemisphere, with a radius of curvature
$R$ and a diameter $2A$, as schemed in Fig.~\ref{spatula}-I. Here, we
assume that a thin film of secretion (thickness $H$) 
initially wets the spatula. 
The duration of the attachment process of the spatula onto the substrate, and
hence of the insect pads, can then be estimated from the above
rheological measurements.

As a spatula approaches the substrate, the contact region of radius
$a$ rapidly expands laterally. 
We shall now estimate the capillary force between the spatula at the setal
tip and a smooth substrate in the case of partial contact ($a<A$).
Immediately outside the contact region of radius $a$, 
the surface of the spatula
is expected to depart from the substrate with a weak slope $\theta$,
with $\theta\simeq a/R$ (Figure~\ref{spatula}-II). 
Considering that there is very little secretion at the tip,
the meniscus width $\Delta$ is much smaller than $a$. 
The meniscus height (which is also twice its
radius of curvature $r$) is then typically equal to
$h\simeq\theta\,\Delta$. 
Integrating the capillary (Laplace) pressure
$-\gamma/r=-2\gamma/h$ over the contact surface area between the
meniscus and the spatula, which is approximately $2\pi\,a\,\Delta$,
provides the capillary force $F_{\rm capil}^{a<A}\simeq
4\pi\,R\,\gamma$. Interestingly, this capillary force turns out to be
independent of both the radius $a$ of the contact and of the amount of
liquid in the meniscus. Note that depending on the spatula curvature and rigidity, the secretion surface
tension and the force applied by the insect, the contact -- of final
radius $a_{\rm fin}$ -- remains
either partial ($a_{\rm fin}<A$) or total
($a_{\rm fin}=A$). In both cases, the dynamics of the
attachment/detachment process remains identical. 

Let us then focus on the attachment dynamics. The viscosity
$\eta$ extracted from the above Brownian motion measurements now
allows us to derive a prediction for the duration of the
attachment/detachment process of a spatula on a substrate. As the
spatula approaches the substrate, the contact
radius widens at velocity
$\dot{a}$. 
The intersection of the film of thickness $H$
deposited on the tip radius of curvature $R$
and the planar substrate
is a disk (radius $a_{\rm in} \simeq \sqrt{HR}$)
which approximately represents the initial contact area. 
The capillary force thus pulls the surfaces together at velocity
$v_z=\dot{a}\theta$, yielding a power $P_{\rm capil}=F_{\rm
capil}\,v_z$, and causes the fluid in the meniscus to move
outwards with an average velocity $\dot{a}>0$, as schemed in
Figure~\ref{spatula}-III.
The zero-velocity condition at the substrate and the spatula surfaces
implies a dissipation of order $\eta(\dot{a}/h(x))^2$ per unit volume,
where the local gap reads $h(x)=\theta x$, with $x$ the horizontal position
along the meniscus width. The total power dissipated in the meniscus
is thus approximately $P_{\rm viscous} \approx\int\eta(\dot{a}/\theta
x)^2\,2\pi a\,\theta x\,{\rm d}x \approx \eta\dot{a}^2 a / \theta$. 
Note that the above calculation
of the dissipation in the meniscus
overlooks the detailed shape of the spatula 
as deformed dynamically by the meniscus
(to the best of our knowledge, 
such a calculation is not currently available).
It is likely that the spatula surface 
does not locally make a sharp (although small)
angle with the substrate,
and that the true dissipated power
is in fact larger than the above estimate.

Equating $P_{\rm viscous}$ and $P_{\rm capil}$, we find that the
contact widens at velocity $\dot{a}\approx a\gamma/R\eta$. The
corresponding time needed to establish the equilibrium contact reads
$T = { \int^{a_{\rm fin}}_{a_{\rm in}}} (1/\dot{a})\, da = \eta
R/\gamma \log({a_{\rm fin}/a_{\rm in}})$. Considering that the
quantity $\log({a_{\rm fin}/a_{\rm in}})$ remains of order unity
($a_{\rm fin}$ not so different from $a_{\rm in}$), the duration of
the attachement process can be estimated to scale as $T\approx\eta
R/\gamma$. Using our rheological measurements $\eta= 10^{-1} $ Pa.s,
and assuming $\gamma \approx 10^{-2}$ N.m$^{-1}$ and $R\approx
1\,\mu{\rm m}$, we obtain $T \approx 10$ ms.

This represents the duration of spontaneous contact formation,
without any active effort from the insect:
$F_{\rm insect}^{\rm attach}=0=F_{\rm capil}-F_{\rm viscous}^{\rm attach}$,
where $F_{\rm viscous}$ is the viscous resistance
related to the dissipated power through 
$P_{\rm viscous}=v_z\,F_{\rm viscous}$.
To achieve detachment, the insect must exert
a tensile force greater than the capillary force:
$F_{\rm insect}^{\rm detach}=F_{\rm capil}+F_{\rm viscous}^{\rm detach}>F_{\rm capil}$.
For detachment to occur at the same rate as attachment,
which seems reasonable, it is required that
$F_{\rm viscous}^{\rm detach}\simeq F_{\rm viscous}^{\rm attach}$,
hence $F_{\rm viscous}^{\rm detach}\simeq F_{\rm capil}$.
As a result, during detachment,
the insect must exert a tensile force
typically equal to $2\,F_{\rm capil}$.

The present picture thus predicts,
on the basis of the measured viscosity,
a total duration $T$ of the contact (attachment {\em and} detachment)
and thus a pace rate,
as well as a detachment force 
$F_{\rm insect}^{\rm detach}\simeq 2\,F_{\rm viscous}$.
It turns out that the estimate $T\simeq 10\,{\rm ms}$
obtained for the contact duration
is compatible with the duration
observed in videorecordings of walking beetles (personal observations).

Concerning detachment, moderate viscous dissipation
is preferable for the insect,
but perhaps not crucial: we speculate that beetles,
like geckoes~\cite{autumn2000} and flies \cite{niederegger2003},
adopt a leg motion suitable for easing detachment significantly.
This may imply bending or rolling of the spatulae in such a way that the
local curvature becomes stronger at the contact with the substrate.

\section{Discussion and conclusion}

Our passive microrheology nanoliter procedure
is innovative in two respects:
{\em (i)} the material under investigation
is initially available as an assembly
of 1 femtoliter droplets deposited on a glass slide,
{\em (ii)} the 0.1\,nl collected volume 
used to perform the microrheological measurement
was significantly less than the usual 1\,$\mu$l.

These measurements show that the secretion behaves as a purely viscous
fluid of high viscosity over the range of frequencies
investigated. 
This the first experimental indication that beetle locomotion
might not need a specifically visco-elastic behaviour.
Let us emphasize, however, that due to the resolution ($ 30 $nm)
of our particle tracking setup,
we could not explore frequencies above $15\,$Hz,
which is slightly below the beetle pace rate (around $100\,$Hz).
It is therefore not to be excluded
that the secretion response should depart 
from a purely viscous behaviour at higher frequencies.
In future studies, we intend to extend our measurements
to the secretions of other animals
over a wider range of frequencies.

Our model provides expression $T\approx\eta R/\gamma$ 
for the duration of the pad/substrate contact formation.
It implies that the secretion viscosity $\eta$
sets an upper bound of the order of $1/T$
for the insect's pace rate.
This formula is only an estimate 
since {\em (i)} the exact spatula geometry and dimensions
are not precisely known and vary
within one single animal~\cite{Voigt2008},
{\em (ii)} numerical prefactors were omitted.
Nevertheless, using this expression with the measured viscosity
provides a timescale for attachment
which is compatible with live observations.
The high secretion viscosity should therefore 
not prevent the insect from forming good contacts.

One might wonder, however, 
why the secretion viscosity is so high
(100 times the water viscosity).
A lower viscosity would ease the insect locomotion
and speed up its pace.
It is not unreasonable to imagine
that high viscosity, and correspondingly high molecular weight,
ensures slow evaporation ---
a crucial issue at such small length scales.
Additionally, a high viscosity ensures that the adhesion 
is robust under unexpected and fast varying conditions.

\section*{Appendix}

\subsection*{Insects and preparation of the footprints}Adult beetles {\it
L. decemlineata} Say (Chrysomelidae) were collected on various species
of annual plants from the family Solanaceae in the Botanical Garden of
Dresden University of Technology, Germany. The tarsi were cut off the
body using small scissors and used to prepare footprints on the
substrate by pressing ventral side of tarsi against clean glass
slide. A pressure was applied on the foot in order to have a strong
contact with the glass slide and to make the secretion go out from the
pore channels. At the same time, the leg is rubbed against the slide, which results
in a strong shear that is able to deposit the secretion on the
surface. It produces a collection of small droplets of various sizes,
from a few microns to ten microns for the largest ones (Figure
\ref{aspiration}). In all cases, the droplets were spread, their
height never exceeding a few microns. Interestingly, evaporation of
the secretions was not observed within the experiment time.

\subsection*{Particle tracking algorithm}Two dimensional analysis of the
particle motion was performed using a home-made algorithm, implemented
as an ImageJ plugin \cite{oc}. A classical cross-correlation method
was used to determine the beads displacements. In order to achieve a
sub-pixel spatial resolution, the computed correlation was
interpolated using a parabolo\"{\i}d approximation. With this
algorithm, several beads could be tracked at the same time.
	
\subsection*{Mean-squared displacement analysis}Passive, or
thermally driven, microrheology is based upon the Generalized Einstein
Relation (GER), valid in a visco-elastic stationary medium in thermal
equilibrium at temperature $T$. It relates the frequency-dependent
mean-squared displacement of the diffusing particle to its
frequency-dependent mobility, according to the relation:
$s^2\langle\Delta \hat x^2(s)\rangle = 2kT\,\hat{\mu}(s)$, where $k$ is the
Boltzmann constant, and $s$ the frequency in the Laplace domain
\cite{mason, waigh, abou08}. As can be seen, measuring the particles mean-squared
displacement at equilibrium leads to the indirect measurement of the particle
mobility. The bulk frequency-dependent viscosity of the fluid can then
be deduced assuming that the Stokes relation $\hat \mu(s)= 1/ 6 \pi R
\hat \eta(s)$
remains valid in the visco-elastic material. In the particular
case of a purely viscous fluid, the mean-squared
displacement increases linearly with time and the GER writes $\langle \Delta
r^2 (t)\rangle= 4Dt$, where $D$ is the diffusion coefficient. The
viscosity of the material is directy deduced from the diffusion
coefficient according to the Stokes-Einstein relation $D= \frac{kT}{6
\pi R \eta_0}$, where $R$ is the bead radius and $T$ the bath
temperature.

\subsubsection*{acknowledgments}
This work was supported by the Programmes
Interdisciplinaires de Recherche from the CNRS
to BA. 
This work as part of the European Science Foundation EUROCORES
Programme FANAS was supported by the German Science Foundation DFG
(contract No GO 995/4-1) and the EC Sixth Framework Programme (contract
No ERAS-CT-2003-980409) to SNG.

\bibliographystyle{plain}       



\end{document}